\documentclass[tinyml]{acmart}
\usepackage{graphicx}
\usepackage{float}
\usepackage{color}
\usepackage{soul}

\AtBeginDocument{%
  \providecommand\BibTeX{{%
    \normalfont B\kern-0.5em{\scshape i\kern-0.25em b}\kern-0.8em\TeX}}}

\setcopyright{rightsretained}
\copyrightyear{2022}
\acmYear{2022}

\graphicspath{ {./Figures/} }
\begin{document}

\title{Distributed On-Sensor Compute System for AR/VR Devices: \\
A Semi-Analytical Simulation Framework for Power Estimation}

\author{Jorge Gomez*, Saavan Patel*, Syed Shakib Sarwar, Ziyun Li, Raffaele Capoccia, Zhao Wang, Reid Pinkham, Andrew Berkovich, Tsung-Hsun Tsai, Barbara De Salvo and Chiao Liu }
\thanks{*Both authors contributed equally to this work}

\email{jtgomez@fb.com}

\affiliation{
  \institution{Meta Reality Labs Research, USA}
 \streetaddress{322 Airport Blvd, USA}
}




\renewcommand{\shortauthors}{Gomez and Patel, et al.}

\begin{abstract}
Augmented Reality/Virtual Reality (AR/VR) glasses are widely foreseen as the next generation computing platform. AR/VR glasses are a complex "system of systems" which must satisfy stringent form factor, computing-, power- and thermal- requirements. In this paper, we will show that a novel distributed on-sensor compute architecture, coupled with new semiconductor technologies (such as dense 3D-IC interconnects and Spin-Transfer Torque Magneto Random Access Memory, STT-MRAM) and, most importantly, a full hardware-software co-optimization are the solutions to achieve attractive and socially acceptable AR/VR glasses. To this end, we developed a semi-analytical simulation framework to estimate the power consumption of novel AR/VR distributed on-sensor computing architectures. The model allows the optimization of the main technological features of the system modules, as well as the computer-vision algorithm partition strategy across the distributed compute architecture. We show that, in the case of the compute-intensive machine learning based Hand Tracking algorithm, the distributed on-sensor compute architecture can reduce the system power consumption compared to a centralized system, with the additional benefits in terms of latency and privacy.

\end{abstract}


\maketitle

\begin{figure*}[h]
  \centering
  \includegraphics[width=\linewidth]{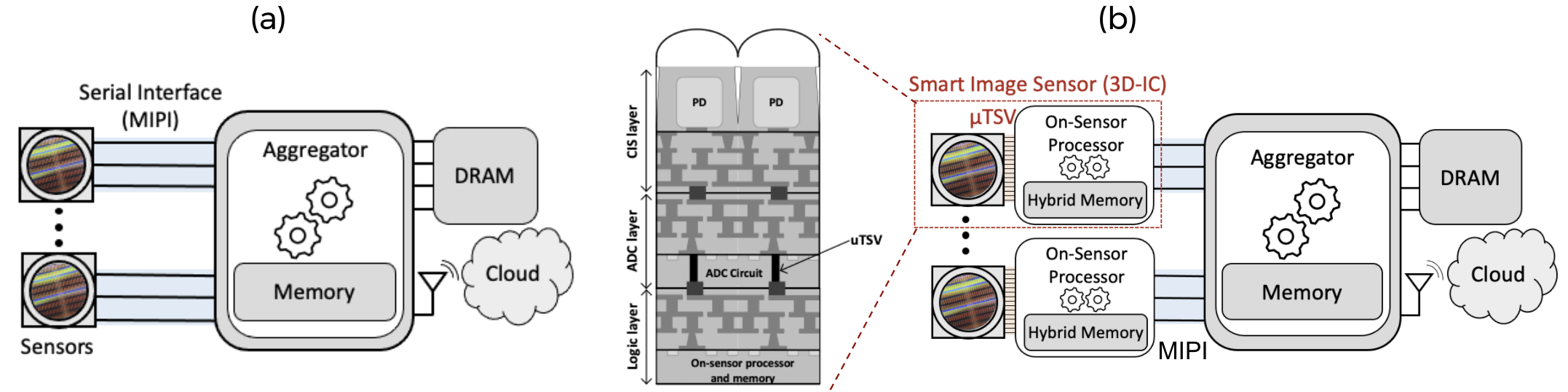}
  \caption{Schematic of (a) Traditional centralized camera-sensor computing architecture, (b) Distributed on-sensor computing architecture \cite{Liu2022}, where each sensor has an integrated on-sensor processor (via $\mathrm{\mu TSV}$ and a hybrid memory hierarchy (i.e. SRAM/STT-MRAM).}
  \label{fig:1}  
\end{figure*}

\section{Introduction}

 \begin{figure*}[h]
  \centering
  \includegraphics[width=\linewidth]{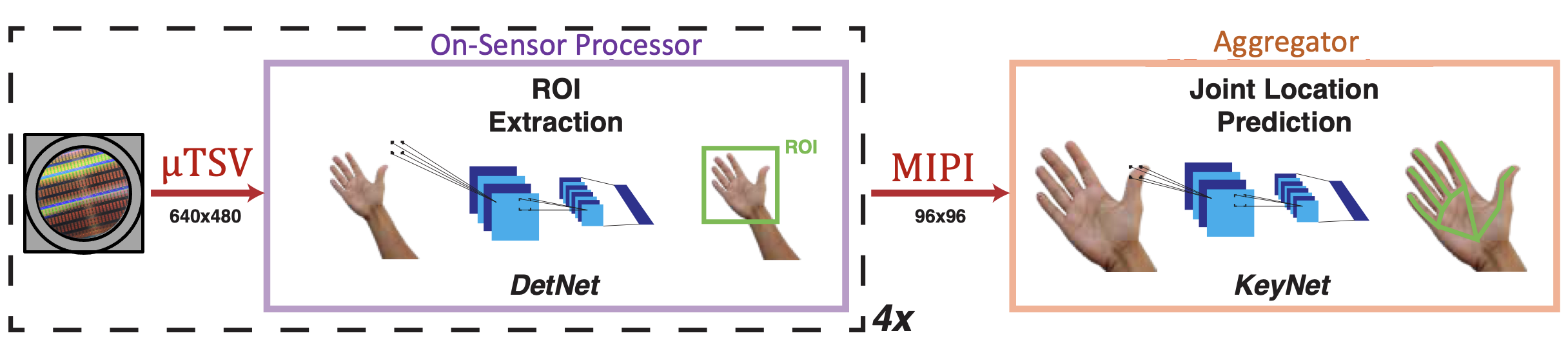}
  \caption{Example of the optimal repartition of the AR/VR Hand Tracking workload \cite{Han2020} on the distributed on-sensor computing architecture. The workload consists in two consecutive NNs (the first for detecting hands, DetNet, and the second to estimate hand keypoint locations, KeyNet). We foresee that the first ML model (DetNet) is deployed on-sensor, while the second model (KeyNet) runs on the aggregator. Only the region of interest (ROI), as extracted from the raw image by the on-sensor processor, is transmitted to the aggregator through the energy-hungry MIPI serial interface.}
  \label{fig:HT}  
\end{figure*}

Mixed Reality, consisting of Virtual Reality (VR) and Augmented Reality (AR) together, will be the next general computing platform, dominating our relationship with the digital world for the next fifty years, much as personal computing has dominated the last fifty \cite{Abrash2021}. AI (Artificial Intelligence) processing is critical for AR/VR glasses, both for input (\textit{e.g.}, camera images, audio) and output (\textit{e.g.}, graphics rendering) \cite{Li2021}. Moreover, the AR/VR glasses has to deliver a responsive, low latency experience while consuming very limited power, both for battery life and to limit temperature rise on the skin for user comfort.
At the system level, a large quantity of data is generated and must be processed in real time to support precise and low latency interaction between the physical world and the virtual world. In traditional AR/VR systems, the processing of the raw input data captured by several image sensors is performed in a central off-sensor edge-processor (named aggregator in Fig. \ref{fig:1} (a)). The new distributed on-sensor compute architecture (shown in Fig. \ref{fig:1} (b)) \cite{Liu2022, Abrash2021, Liu2020} exploits multiple levels of computing, including on-sensor processors (for initial feature extraction and first level of processing) and a nearby aggregator for further processing and final analysis. The distributed compute system allows for minimum data movement with rapid and localized inference on the sensors, where a shallow portion of each Computer-Vision Neural Network (CV NN) model is implemented. This results in significant benefits in terms of communication costs, latency constraints and privacy concerns compared to centralized computing systems. On the other hand, several challenges need to be solved to implement an on-sensor processing units capable of tackling real-time compute/memory intensive workloads in the very small footprint and low power budget of AR/VR image sensors. As illustrated in Fig. \ref{fig:1} (b), innovative 3D
CMOS Image Sensor (CIS) architectures will permit further shrinking of the sensor footprint and introduction of key on-sensor AI functionalities. Complex three-layer wafer stacking technologies are a promising solution for intelligent image sensor fabrication with AI capabilities. Dense and tight 3D-IC technologies, such as micro-Trough Silicon Vias ($\mathrm{\mu TSV}$) and wafer-level hybrid bonding, will enable heterogeneous monolithic integration of camera and processor. There is a clear need for integrating advanced logic as close to the sensors as possible. Moreover, use of hybrid memory hierarchy (including both SRAM and STT-MRAM) in the on-sensor computing architecture will allow for higher density and lower power. 
To quantify the trade-offs of Distributed On-Sensor Compute (DOSC) architectures for AR/VR workloads, a system simulation framework is required. Several accelerator simulation frameworks  have been developed in recent years \cite{Binkert2011, Parashar2019, Pinkham2021}, mostly focused on the compute modules of the system. In this work, we propose a new semi-analytical model that captures the key technological features   of the whole system hardware resources, including cameras, communication links, processors, and memories. Technological parameters used in the simulations for all modules are silicon-based. The compute and memory parameters of the system have been calibrated by using an event-based silicon-calibrated simulator (GVSoC  \cite{ Burrello2021, GreenwavesTechnologies2021}). Finally, for the system modeling demonstration, we focus on a common AR/VR workload:  the Hand Tracking algorithm \cite{Han2020}  (see Fig. \ref{fig:HT}). In the case of a centralized compute system the complete image captured by the cameras is transmitted from the sensors to the aggregator through the MIPI and the full Hand Tracking workload is deployed on the aggregator. On the other hand, on a hierarchical compute system, an optimal repartition of the workload between the on-sensor processor and the aggregator is possible. In the distributed on-sensor compute system, the complete raw image is transmitted from the cameras to the on-sensor processors through the low-energy and high bandwidth $\mathrm{\mu TSV}$ interconnects. The region of interest (ROI) is thus computed on sensor, and only this data is transferred from the sensors to the aggregator through the energy-hungry MIPI interfaces.  As detailed in \cite{Han2020}, the system uses four monochrome cameras.  By means of our semi-analytical modeling, we will show that this optimal algorithm repartition across the novel distributed on-sensor compute architecture, as well as the introduction of new emerging technologies, allow for a significant improvement of the overall system power consumption. 

\section{Semi-Analytical System Modeling}
In order to evaluate the energy efficiency of the distributed on-sensor computing system (Fig \ref{fig:1}.(b)), we consider separately the system key modules, such as: cameras, communication links, on sensor processor and the corresponding on-sensor memory hierarchy, as well as aggregator processor and its memory hierarchy. 
We assume that the total energy per frame of the compute system is simply the addition of the energies of each module, \textit{i.e.}: 

\begin{multline}
  E_{Total} = \sum ^{\#(Cameras)}_{h=0}E_{Ca,h}+ \sum ^{\#(Comm)}_{i=0}E_{Comm,i} + \\
  \sum ^{\#(Comp)}_{j=0}E_{Comp,j} + \sum ^{\#(Mem)}_{k=0} \left( E_{Read/Write,k}+E_{Lk,k} \right)
  \label{eq:1}
\end{multline}

where $E_{Ca,h}$, $E_{Comm,i}$, $E_{Comp,j}$, $E_{Read/Write,k}$ and $E_{Lk,k}$ are the total camera, communication, compute, read, write, and leakage energy per frame (as calculated in eq.\ref{eq:cam}, eq.\ref{eq:com}, eq.\ref{eq:comp}, eq.\ref{eq:acc} and eq.\ref{eq:lk} respectively) for each camera (h), communication link (i), compute processor (j) and memory instance (k) present in the system. 
To calculate the average power ($P_{Avg}$), we simply multiply by the fps at which each module operates, which gives: 

\begin{multline}
  P_{avg} = \sum ^{\#(Cameras)}_{h=0}E_{Ca,h}\times fps_h+ \sum ^{\#( Comm)}_{i=0}E_{Comm,i}\times fps_i + \\ 
 \sum ^{\#(Comp)}_{j=0}E_{Comp,j}\times fps_j + \sum  ^{\#(Mem)}_{k=0} \left( E_{Read/Write,k}+E_{Lk,k} \right) \times fps_k 
\end{multline}

For each module, we investigate the critical parameters that define its energetic behaviour and we derive simple analytical expressions to capture this behavior (as explained in the following subsections). Once we fix the AR/VR workload, the key parameters of the on-sensor and aggregator processor behavior and optimal dataflow across the multi-level memory hierarchy were extracted by means of the GVSoC tool (a light-weight, event-based instruction set simulator \cite{GreenwavesTechnologies2021}) and DORY \cite{Burrello2021}, respectively.

\subsection{Digital Pixel Sensor}

\begin{table}
  \begin{tabular}{c|c|c}
    \toprule
    Camera  & Operation State & Power (mW) \\
    \midrule
                         & Sensing & 15   \\ \cline{2-3}
    DPS & Read Out & 36   \\ \cline{2-3}
                         & Idle & 1.5  \\
  \bottomrule
\end{tabular}
 \newline  \newline
\caption{Values used in simulations for power of the camera in different operating states, based on a custom AR/VR digital-pixel-sensor \cite{Liu2020})}
\label{tab:cam}
\end{table}

We consider the image sensors of the distributed on-sensor compute system being based on a AR/VR custom digital pixel sensor (DPS) technology (detailed elsewhere \cite{Liu2020}). 
The camera analytical model assumes that the sensing and readout energy are proportional to the sensing time, $T_{Sense}$, and the readout time, $T_{Rd}$, respectively. The sensing time includes the exposure and the analog to digital converter (ADC) times. We calculate the energy of the camera as follows:

\begin{equation}
  E_{Ca} = P_{Sense} \times T_{Sense}+P_{Rd} \times T_{Comm} + P_{Off} \times T_{Off}
  \label{eq:cam}
\end{equation}

where $P_{Sense}$ , $P_{Rd}$ and $P_{Off}$  are the sensing, read out and off power respectively (Table \ref{tab:cam}). The sensing time is the sum of the exposure and ADC time ($T_{Sense} = T_{Ex} + T_{ADC}$). $T_{Comm}$ is the read-out time (see later Eq. \ref{eq:Tif}) that depends on the interface between the camera and the compute module. Finally, $T_{Off}$ is the rest of the time which can be simply calculated as follows:

\begin{equation}
  T_{Off} = \frac{1}{fps}-T_{Sense}-T_{Comm}   
\end{equation}

The power values estimated from the DPS (\cite{Liu2020}) are shown in Table \ref{tab:cam}.

\subsection{Communication Links}

Different interfaces are used in the DOSC architecture, as shown in Fig. \ref{fig:1} (b), in particular: $\mathrm{\mu TSV}$ between the image sensor and the on-sensor compute layer, and MIPI between the on-sensor processor and the aggregator. Each interface is characterized by the energy it requires to transmit a Byte of information ($E_{Byte,Comm}$) and by its bandwidth ($BW_{Comm}$). The total energy of the communication link ($E_{Comm}$) is given by the following equation:
\begin{equation}
  E_{Comm} = A_{Size} \times E_{Byte,Comm}
  \label{eq:com}
\end{equation}

where $A_{Size}$ is the size of the data (in Bytes) transmitted trough the interface, which depends on the workload. As shown in Fig. \ref{fig:HT},  $A_{Size}$ can correspond to the raw image (transferred from the camera to the on-sensor processing layer), the Region of Interest (ROI) from the input image, or it could be the Neural Network activation map output (transferred from the on-sensor processing unit
to the aggregator). 
The communication time ($T_{Comm}$) is then equal to: 

\begin{equation}
  T_{Comm} = \frac{A_{Size}}{BW_{Comm}}
  \label{eq:Tif}
\end{equation}

The energy and bandwidth values corresponding to the two interfaces $\mathrm{\mu TSV}$ and MIPI used in our distributed system are given in Table \ref{tab:int} (based on literature data).

\begin{table}

  \begin{tabular}{c|c|c|c}
    \toprule
   \shortstack{Communication \\ Link}  & \shortstack{Energy per \\ Byte (pJ/B)} & \shortstack{Bandwidth \\ (GB/s)}   & Ref.\\
    \midrule
    $\mathrm{\mu TSV}$  & 5 & 100 & \cite{TSV} \\ \hline
    MIPI & 100 & 0.5 &  \cite{MIPI1, Takla2017}  \\
  \bottomrule
\end{tabular}
\newline  \newline
  \caption{Values used in simulations for the energy and bandwidth of the communication links, based on literature data.}
  \label{tab:int}
\end{table}

\subsection{On-Sensor and Aggregator AI Processors and Memories}

\begin{figure}[t]
  \centering
  \includegraphics[width=\linewidth]{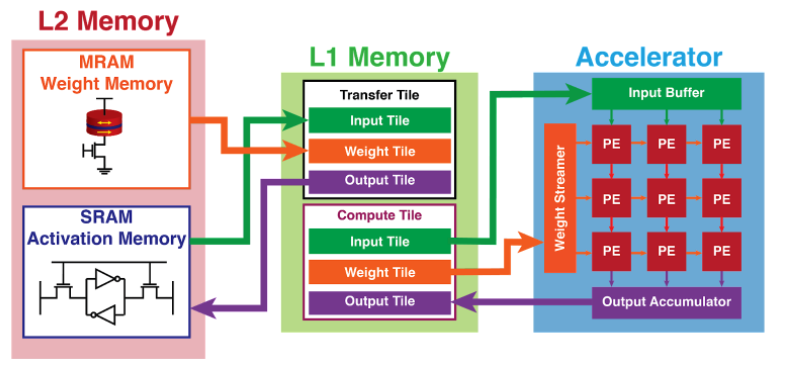}
  \caption{ Memory hierarchy and DNN hardware accelerator used in our simulations. We consider two separate L2 memories, an activation memory (implemented in SRAM), and a read-dominated weight memory (implemented in SRAM or MRAM). We optimized the tiling strategy and dataflow among the different memory levels for our workloads using DORY tiling engine  \cite{Burrello2021}.}
  \label{fig:2}  
\end{figure}

\begin{figure}[ht]
  \centering
  \includegraphics[width=\linewidth]{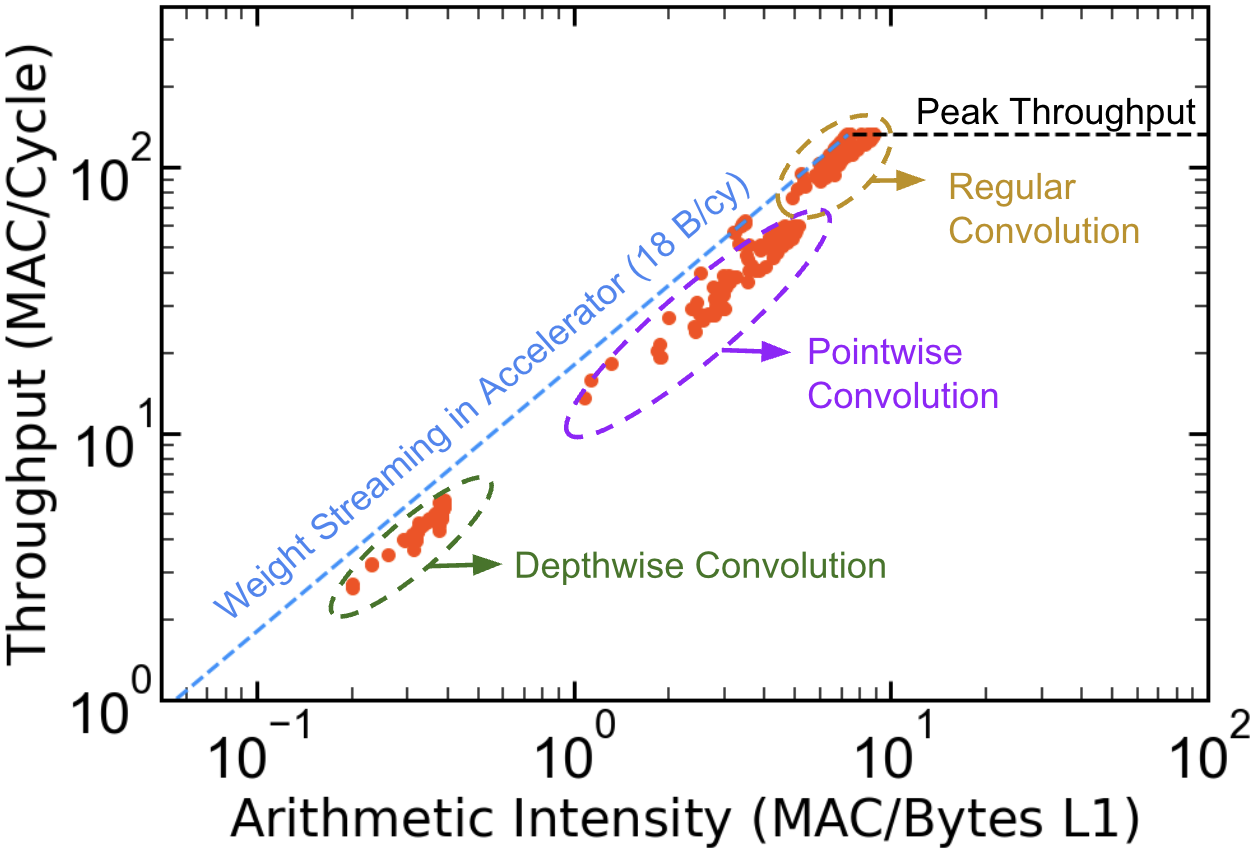}
  \caption{ Roofline plot generated using GVSoC when different NN layers are deployed on the RBE accelerator and memories. It appears that several layers are memory-bounded by weight streaming in the accelerator.}
  \label{fig:3}  
\end{figure}

The computing architecture used for the on-sensor processing and aggregator is based on the PULP computing platform \cite{Pullini2019, GreenwavesTechnologies2021, Valente2021}. In this work, we assume that the AR/VR workload computation is completely done in the DNN hardware accelerator, a Reconfigurable Binary Engine (RBE) \cite{RBE}. The RBE has a maximum throughput of 133 MAC/Cycle, running on 8-bit precision weights and activations. As shown in Fig. \ref{fig:2}, the simulated memory hierarchy includes two levels: a small Level 1 memory and two larger Level 2 memories (activation memory and weight memory). We separate L2 activation from L2 weight memory, in order to explore the impact of using STT-MRAM as L2 weight memory.

In our semi-analytical model, the compute energy per frame is given by:

\begin{equation}
  E_{Comp} = \#(MACs) \times E_{MAC}
  \label{eq:comp}
\end{equation}

 where $\#(MACs)$ is the total number of Multiply Accumulate (MAC) operations for processing one frame and $E_{MAC}$ is the energy per MAC operation, depending on the specific technology node. 
 
The access energy for the different memory levels (L1 or L2) is then computed as follows:

\begin{equation}
  E_{Read/Write} = \#(Read) \times E_{Byte,Read} +  \#(Write) \times E_{Byte,Write}
  \label{eq:acc}
\end{equation}
 
 where $\#(Read/Write)$ is the total number of read or write (for an specific memory level) and $E_{Byte,Read/Write}$ is the read or write energy per byte (for the specific memory level and the specific technology node).

To calculate the impact of the memory leakage, we first need to calculate the amount of time that the memory is in On-state, in Retention-state, or Off-state. To this scope, we use the following equation:
 
\begin{equation}
  T_{Processing} = \sum_{j=0}^{\#(layers)} \frac{\#(MAC_j)}{(MAC/cycle)_j} \times \frac{1}{f_{clk}}   
\end{equation}

where  $\# MAC_j$ is the number of MACs of layer j, $(MAC/cycle)_j$ is the throughput that the system has at layer j and $f_{Clk}$ is the clock frequency. The processing time is the time that the memory will be in On-state. The rest of the time the memory can be in Retention- or Off-state, depending on the memory type. We can simply calculate the idle time by using the following equation: 

\begin{equation}
  T_{idle} = \frac{1}{fps}-T_{Processing}  
\end{equation}

Finally, the total memory leakage energy per frame for each memory level will be given by the following equation: 

\begin{equation}
  E_{Lk} =  T_{Processing} \times Lk_{On} + T_{idle} \times Lk_{Ret/Off}
  \label{eq:lk}
\end{equation}

where $Lk_{On}$ and $Lk_{Ret/Off}$ are the leakage power in On, Retention or Off state respectively, depending on the specific technology.

The energy and power values used in simulations for the elementary MAC operation and the memory read/write access or leakage (\textit{i.e.} $E_{MAC}$, $E_{Byte, Read/Write}$, $Lk_{On}$ and $Lk_{Ret/Off}$) are extracted from post-synthesis simulations and memory compilers, respectively. In this work, we used 7nm and 16nm logic process nodes and libraries from leading foundries. STT-MRAM values correspond to values from test-vehicles fabricated in 16nm logic technology \cite{Guedj2021}. 

On the other hand, the number of operations, the processing efficiency, the memory operations counts and accelerator performance ($\#MAC$, $\#MAC_j$, $(MAC/cycle)_j$, and $\#(Read/Write)$) are obtained by using the GVSoC/Dory/Nemo toolchain \cite{Burrello2021, GreenwavesTechnologies2021} to simulate the deployment of our AR/VR workload on the computing platforms. GVSoC allows us to capture the processing performance depending on the layer configuration and the arithmetic intensity of the NN layer (on each memory level). This is especially important when complex memory hierarchies are considered with different capacities, bandwidths and non-symmetric read-write performances.
To deploy the workload across the different memory levels, we modified  DORY tiling engine \cite{Burrello2021} to work with the RBE accelerator. 

 Using GVSoC we also characterized several representative layers of our workloads that include among others Regular, Depthwise separable, and Pointwise convolutions with varying channel dimension and spatial fields. Using these simulations we obtained the roofline plot for the RBE accelerator shown in Fig \ref{fig:3}. The roofline plot allows us to understand the accelerator performance in relation to memory bandwidth constraints. From the plot, we can see that layer performance is almost completely bounded by the weight streaming in the accelerator. The RBE demonstrates close to peak performance on full convolutional benchmarks, with diminishing performance for pointwise kernels, and even further decrease when doing depthwise kernels. We note that the RBE has been designed and optimized for a particular architecture, memory bandwidth, and set of workloads. Because of this, the architecture is most efficient for certain network types as compared to others, as shown by the high throughput of convolution kernels compared to pointwise and depthwise kernels. 

\begin{figure*}[h]
  \centering
  \includegraphics[width=\linewidth]{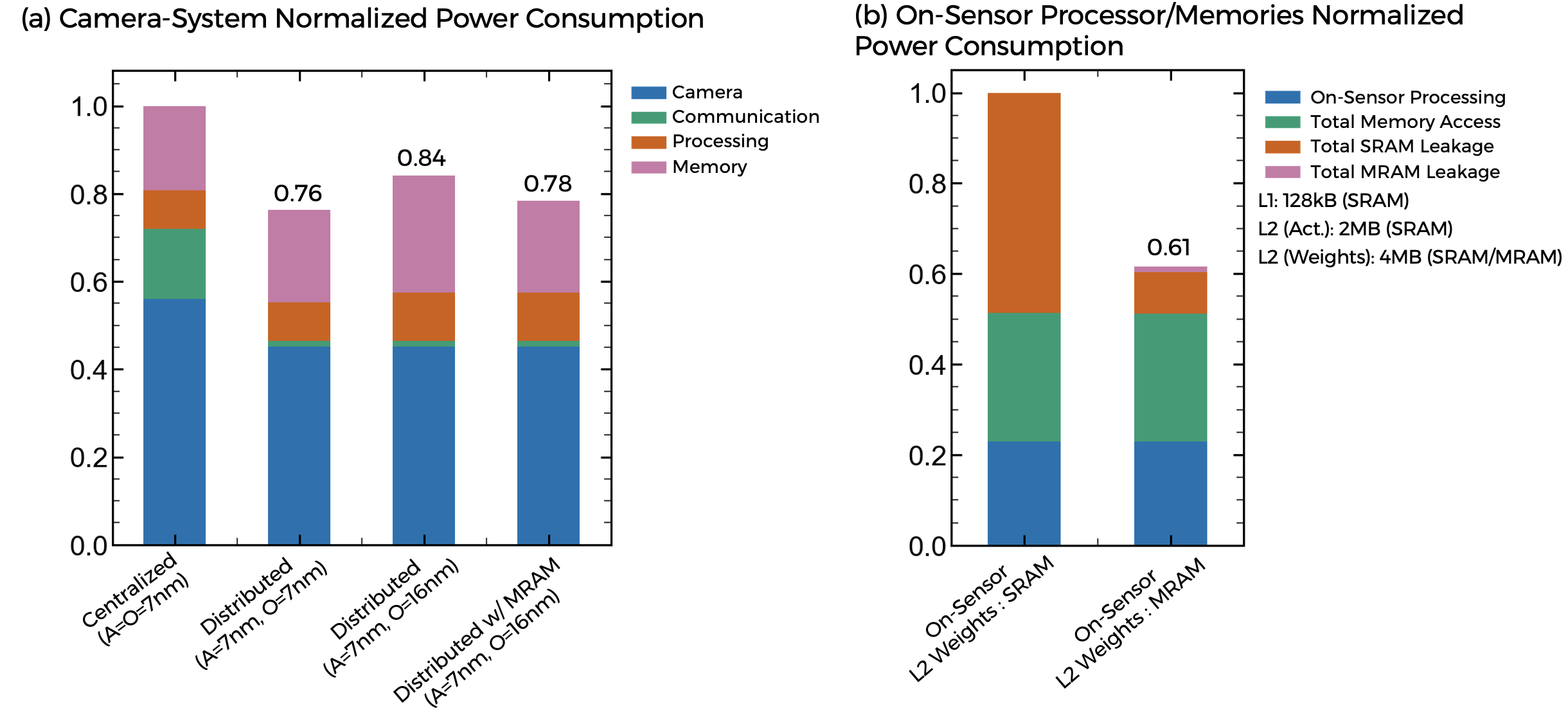}
  \caption{(a) Overall power comparison of centralized and distributed camera compute systems, showing the portion of different components when the AR/VR hand-tracking workload is deployed on the system. The power values are normalized to the case of a centralized system with aggregator in 7nm. Simulations assuming different technology nodes for the aggregator (A) and the on-sensor processors (O) are shown. (b) Power consumption of the on-sensor processor and corresponding memory system, assuming a pure SRAM on-sensor memory hierarchy or a hybrid on-sensor memory hierarchy (SRAM as both L2 activation memory and L1 cache and STT-MRAM as L2 weight memory). Here, we assume that the on-sensor processor is run at 10fps. Power values are normalized to the case of a pure SRAM memory hierarchy. Simulations are based on 16nm technology values. }
  \label{fig:5}  
\end{figure*}

Now that we have all the blocks to calculate the system energy per frame and the average power, we will use our semi-analytical simulation model to explore the Hand Tracking workload. 

\section{Distributed On-Sensor System Simulations for Hand-Tracking}

The semi-analytical model was used to compare the energy efficiency of the DOSC system versus a traditional centralized compute system, while deploying a Hand-Tracking (HT) workload for AR/VR experience. As mentioned previously, the HT NN consists in two consecutive Machine Learning algorithms (Fig. \ref{fig:HT}). The first ML model (DetNet) is used to define the region of interest (ROI), while the second model (KeyNet) is used to detect the joint locations and hand pose. Note that the two ML models, on sensor and in the aggregator, can run at different frames-per-second giving us an additional knob for power optimization. As suggested to improve energy efficiency in \cite{Han2020}, the DetNet model is not run at every frame as the same ROI can be used for multiple frames. In our simulations, we assume that the on-sensor compute capability and corresponding memory size to be one fourth of the aggregators. The L2 weight memories were sized to hold the full weights of the models. The weight memories’ operations are mainly read-dominated. On the contrary, the activation memories need to support both read and write operations during the inference. 
This network has many features which help demonstrate it as a representative workload for our applications. Namely, it is a common and necessary AR/VR application, it has a natural partition point to demonstrate how computation can be split between on-sensor and off-sensor, and it is sufficiently computationally intensive to strain many current systems. 

As shown in Fig. \ref{fig:5}, our simulation results demonstrate that the cameras and MIPIs dominate the power dissipation of the centralized compute system. On the other hand, when we migrate from a centralized to a distributed system, results show a significant system power reduction (24\%, as shown in Fig. \ref{fig:5} (a)). Even when the on-sensor processor technology node is less advanced than the aggregator’s (16nm rather than 7nm), there is still a significant power reduction (16\%). The power gain is mainly due to the decreased usage of the energy-hungry serial interface (MIPI) thanks to the on-sensor processing and subsequent feature compression. Additionally, the image sensor power is also reduced. This is because the wider data bandwidth of the $\mathrm{\mu TSV}$ interconnects (between the camera and the on-sensor processor) compared to MIPI allows us to reduce the sensor digital data read-out time and consequently to extend the sensor low-power standby mode. However, it also appears that the total memory energy consumption slightly increases in the distributed computing system due to the duplication of the weight storage memory in each sensor, increasing the total memory size of the system and thus the leakage energy contribution. Finally, Fig. \ref{fig:5} (b) shows that a hybrid on-sensor memory architecture (STT-MRAM for weight storage and SRAM for activation) could reduce the overall on-sensor power consumption by 39\%  thanks to STT-MRAM’s negligible leakage. Moreover, the use of STT-MRAM also allows us to improve the overall on-sensor memory’s form factor, as STT-MRAM features approximately 2x higher memory density than SRAM \cite{Guedj2021}.

\section{Conclusion}

In this paper, we proposed a semi-analytical simulation framework suitable for optimization of AR/VR distributed on-sensor computing architectures. This model allowed us to study simultaneously the impact of the main technological features of the system modules (cameras, interfaces, accelerators, memories) on the overall system power consumption, as well as the impact of the partitioning of the deployed AR/VR CV NNs across the different resources. When an AR/VR Hand Tracking optimized workload is  partitioned over the distributed system, the model predicts significant power savings compared to traditional centralized systems, thanks to:

\begin{enumerate}
\item Improved communication power, as only high-level data representation is transferred from the sensors to the aggregator through the energy hungry MIPI interfaces, while the first level of processing is performed on sensor.

\item Reduced camera read-out time (and consequently reduced camera power consumption), thanks to monolithic integration of camera and on-sensor processor through high bandwidth/low energy $\mathrm{\mu TSV}$ interconnects

\item Use of STT-MRAM as L2 weight memory on sensor, allowing for 2x higher memory density (thus smaller area) and improved power consumption, thanks to the negligible leakage of STT-MRAM compared to SRAM.

\end{enumerate}
Finally, a significant reduction in the system power remains when the on-sensor processor is implemented in an older technology node than the aggregator’s. However, the older technology node will penalize the on-sensor processor form factor.

\begin{acks}
The authors would like to thank Francesco Conti, Davide Rossi, Alessio Burrello, Nazareno Bruschi, Arpan Prasad, Luca Benini of the Integrated Systems Laboratory (IIS) of ETH Zürich for their valuable advice and support with the DORY/NEMO compilation tool chain and computing system architecture definition.
\end{acks}

\bibliographystyle{ACM-Reference-Format}

\bibliography{References/OnSensorCompute.bib}

\end{document}